\documentclass[aps,pre,twocolumn,showpacs,longbibliography]{revtex4-1}
\usepackage{bbding}
\usepackage{amsmath}
\usepackage{amsthm}
\usepackage{graphicx}
\usepackage{subfigure}
\usepackage{amssymb}
\usepackage{graphicx}
\usepackage{dcolumn}
\usepackage{bm}
\usepackage{float}
\usepackage[colorlinks,linkcolor=blue,citecolor=blue,urlcolor=blue,hyperindex,breaklinks]{hyperref}

\begin{document}

\title{Ground-state cooling of an magnomechanical resonator induced by
magnetic damping}
\author{Ming-Song Ding$^1$}
\author{Li Zheng$^2$}
\author{Chong Li$^1$}

\affiliation{$^1$School of Physics, Dalian University of Technology, Dalian 116024, China\\
$^2$Science and Engineering College, Dalian Polytechnic University, Dalian
116034, China}

\affiliation{$^1$School of Physics, Dalian University of Technology, Dalian 116024, China\\
$^2$Science and Engineering College, Dalian Polytechnic University, Dalian
116034, China}
\email{lichong@dlut.edu.cn}
\date{\today }

\begin{abstract}
Quantum manipulation of mechanical resonators has been widely applied in
fundamental physics and quantum information processing. Among them, cooling
the mechanical system to its quantum ground state is regarded as a key step.
In this work, we propose a scheme which one can realize ground-state cooling
of resonator in a cavity magnomechanical system. The system consists of a
microwave cavity and a small ferromagnetic sphere, in which phonon-magnon
coupling and cavity photon-magnon coupling can be achieved via
magnetostrictive interaction and magnetic dipole interaction, respectively.
After adiabatically eliminating the cavity mode, an effective Hamiltonian
which consists of magnon and mechanical modes is obtained. Within
experimentally feasible parameters, we demonstrate that the ground-state
cooling of the magnomechanical resonator can be achieved by extra magnetic
damping. Unlike optomechanical cooling, magnomechanical interaction is
utilized to realize the cooling of resonators. We further illustrate the
ground-state cooling can be effectively controlled by the external magnetic
field.
\end{abstract}

\maketitle

\section{Introduction}

In recent years, the ferrimagnetic systems has attracted considerable
interest, which can realize the strong light-matter interactions. Among
them, the cavity-magnon systems developed from cavity quantum
electrodynamics (QED) systems have been theoretically proposed and
experimentally demonstrated in the past few years \cite%
{p0,p1,p2,p3,p4,i1,i2,i3,i4,i5,i51,i52,i53,g9}. A highly polished
single-crystal yttrium iron garnet (YIG) sphere is placed in a 3-dimensional
rectangular microwave cavity. Due to YIG sphere's high spin density and low
damping rate, the Kittel mode \cite{k15} (the ferromagnetic resonance mode)
in it can strongly\cite{i3,i6,i7} and even possibly ultrastrong \cite{i8}
couple to the microwave cavity photons. Compared with previous quantum
systems, the cavity-magnon system has the merits of high tunability and good
coherence. This system has shown its extensive application prospects,
including the bistability of cavity magnon polaritons \cite{k6}, the
higher-order exceptional point \cite{k211}, magnon Kerr effect in strong
coupling regime \cite{k2}, the light transmission \cite{k23} and other
researches \cite{k24,k25,k5,k7,k21}.

Cavity optomechanics is a rapidly developing research area for exploring the
radiation-pressure-mediated interaction between photons and mechanical
systems \cite{k00,k000,k001,k002,k003,k004,k005,k006,k007}. In the view of
these important results, it is natural to investigate the magnon-phonon
interaction in YIG spheres and their quantum characteristics. So far, some
approaches on cavity magnomechanical systems have been proposed \cite%
{k0,k1,k151,k16,k26}, where the interaction between magnons and phonons can
be achieved by the magnetostrictive force (radiation-pressure-like). The
varying magnetization caused by the excitation of the magnons in the YIG
sphere results in the geometric deformation of the surface, introducing the
coupling between magnon mode and phonon mode. Generally speaking, the
magnomechanical state-swap interaction is much small in experiments and can
be neglected. However, it can be significantly enhanced by driving the
magnon mode with a strong red-detuned microwave field \cite{k6,k0}.

The cooling of the systems to theirs ground states is a prerequisite for
observing the signature of quantum effects in mechanical systems \cite%
{k27,k28,k29,k30,k31,k32,k33,k34,k35,k36,g5}. Accordingly, preparing
mechanical quantum states free of thermal noise is a crucial goal in various
mechanical systems. Among them, the cavity magnomechanical system has
attracted much interest because of its intrinsic great tunability, low loss,
and promising integration with electromechanical elements. In addition, the
cavity magnomechanical system provides a promising platform for the study of
macroscopic quantum phenomena.

Here, we explore theoretically the ground-state cooling of the
magnomechanical resonator with experimentally reachable parameters. The
system consists of a YIG sphere placed in a low-Q microwave cavity, and an
uniform external bias magnetic field. Base on the highly dissipative cavity
mode, the system is effectively transformed into a two-mode system, which is
composed of a magnon mode and a mechanical mode. Then we give the expression
of the final phonon number in the full quantum theory, it illustrates that
the magnon mode can be utilized to cool the resonator to its ground-state.
In order to achieve better cooling effect, a drive magnetic field is
introduced to enhance the magnon-phonon coupling. In addition, the ground
state cooling can be well controlled by adjusting the external magnetic
field without changing other parameters, which provides an additional degree
of freedom.

The structure of the paper is as follows. In Sec. II, we frist establish a
physical model and give the linearized Hamiltonian. Then, by assuming the
cavity mode is highly dissipative, the effective Hamiltonian by
adiabatically eliminating optical mode is given. We also show the equations
of motion. In Sec. III, we give the analytic expression of extra magnetic
damping for the magnomechanical resonator, then the final phonon number is
studied. We also find the ground-state cooling of an magnomechanical
resonator can be achieved with the experimentally feasible parameters. The
effects of external magnetic field on the cooling are also studied. Finally,
we make a conclusion based on the results obtained in sec. IV.

\section{Model and effective Hamiltonian}

We utilize a hybrid system as shown in Fig. 1(a), which consists a microwave
cavity and a small sphere (a highly polished single-crystal YIG sphere of
diameter $1mm$ is used in \cite{k2}). The YIG sphere is placed near the
maximum microwave magnetic field of the cavity mode, and we add an
adjustable external magnetic field $H$ in $z$-axis, which establishes the
magnon-photon coupling \cite{k2,k6}, and the rate of coupling can be tuned
by the position of YIG sphere in the cavity. The adjusting range of bias
magnetic field $H$ is between $0$ and $1T$ \cite{k2}.

Because of the magnetostrictive effect, YIG sphere can be used as an
excellent mechanical resonator. Therefore, the term of coupling between
magnons and phonons can be introduced into Hamiltonian of our system as
mentioned in \cite{k1}. Based on it, we have the vibrational mode (phonons)
of the sphere.

There are three modes in this system: cavity photon mode, magnon mode and
phonon mode. The equivalent coupled-harmonic-resonator model is given in
Fig. 1(b), and we assume that the cavity mode is highly dissipative. Here, a
microwave source is used to directly drive the magnon mode, therefore, the
magnomechanical coupling can be enhanced \cite{k2,k0}. Moreover, the size of
the sphere we considered is much smaller than the wavelength of the
microwave field. Accordingly, the interaction between cavity microwave
photons and phonons due to the effect of radiation pressure is neglected.
After making a frame rotating at the drive frequency $\omega _{d}$ and using
the rotating-wave approximation, the total Hamiltonian of hybrid system can
be written as ($\hbar =1$)

\begin{eqnarray}
H_{total} &=&-\Delta _{a}a^{\dagger }a-\Delta _{m}m^{\dagger }m+\omega
_{b}b^{\dagger }b+  \notag \\
&&g_{ma}(a^{\dagger }m+m^{\dagger }a)+g_{mb}m^{\dagger }m(b+b^{\dagger })
\label{eq01} \\
&&+i(\varepsilon _{d}m^{\dagger }-\varepsilon _{d}^{\ast }m),  \notag
\end{eqnarray}%
where $\Delta _{a}=\omega _{d}-\omega _{a}$ and $\Delta _{a}=\omega
_{d}-\omega _{m}$ are the detunings, $\omega _{b}$ denotes the resonance
frequency of the mechanical mode. A uniform magnon mode in the YIG sphere at
frequency $\omega _{m}=\gamma _{g}H$, where $\gamma _{g}/2\pi =28GHz/T$ is
gyromagnetic ratio, and we set $\omega _{m}$ at the frequency of Kittel mode
\cite{k15} (uniformly precessing mode), which can strongly couple to the
microwave cavity photons leading to cavity polaritons. $a(a^{\dagger })$, $%
m(m^{\dagger })$ and $b^{\dagger }(b)$ are the annihilation(creation)
operators of the cavity mode, magnon mode and mechanical mode, respectively.
In addition, $g_{ma}$ and $g_{mb}$ are the coupling rates of the
magnon-cavity interaction and magnon-phonon interaction. $i(\varepsilon
_{d}m^{\dagger }-\varepsilon _{d}^{\ast }m)$ is the Hamiltonian which
describes the external driving of the magnon mode.

As we know, $g_{mb}$ is much weak in the experiment \cite{k1}. The
magnetostrictive coupling strength is determined by the mode overlap between
the uniform magnon mode and the phonon mode. J. Q. You $et$ $al.$ designed
an experimental setup, where the YIG sphere can be directly driven by a
superconducting microwave line which is connected to the external port of
the cavity \cite{k2,k6}. Rabi frequency $\varepsilon _{d}=\frac{\sqrt{5}}{4}%
\gamma _{g}\sqrt{M}B_{0}$ (under the assumption of the low-lying
excitations) stands for the coupling strength of the drive magnetic field
\cite{k0}. The amplitude and frequency are $B_{0}$ and $\omega _{d}$
respectively, the total number of spins $M=\rho V$, where $V$ is the volume
of the sphere. Furthermore, $\rho =$ $4.22\times 10^{27}m^{-3}$ is the spin
density of the YIG sphere.

The dynamics of the system can be linearized, through a series of
calculations, we have the linearized Hamiltonian (see Appendix A)

\begin{eqnarray}
H_{lin} &=&-\Delta _{a}a^{\dagger }a-\tilde{\Delta}_{m}m^{\dagger }m+\omega
_{b}b^{\dagger }b+  \label{eq02} \\
&&(Gm^{\dagger }+G^{\ast }m)(b+b^{\dagger })+(g_{ma}m^{\dagger
}a+g_{ma}^{\ast }ma^{\dagger }),  \notag
\end{eqnarray}%
where $\tilde{\Delta}_{m}=\Delta _{m}-g_{mb}(\beta +\beta ^{\ast })$ is the
modified detuning of the magnon mode, For the parameters we considered here,
$g_{mb}(\beta +\beta ^{\ast })\ll \Delta _{m}$, so we can approximately have
$\tilde{\Delta}_{m}\approx \Delta _{m}$. $G=\eta g_{mb}$ can be regarded as
the coherent-driving-enhanced magnomechanical coupling strength with $\eta $
the average magnetic field of magnon mode. We have $\beta
=-ig_{mb}\left\vert \eta \right\vert ^{2}/(i\omega _{b}+\gamma _{b})$ by
solving steady-state Langevin equations, and $\eta $ is given by
\begin{equation}
\eta =\frac{\varepsilon _{d}(-i\Delta _{a}+\kappa _{a})}{g_{ma}^{2}+(-i%
\Delta_{m}+\kappa _{m})(-i\Delta _{a}+\kappa _{a})},  \label{eq021}
\end{equation}%
where $\kappa _{a},\kappa _{m}$ and $\gamma _{b}$ are the losses of
microwave cavity mode, magnon mode and mechanical mode, respectively.
Because $\eta $ is affected by the driving field, We can enhance $G$ by
tuning the external driving field $\varepsilon _{d}$. Note that the
nonlinearity in $\Delta_{m}$ comes from $\left\vert \eta \right\vert ^{2}$
intrinsically.

\begin{figure}[tbp]
\centering\includegraphics[width=7cm]{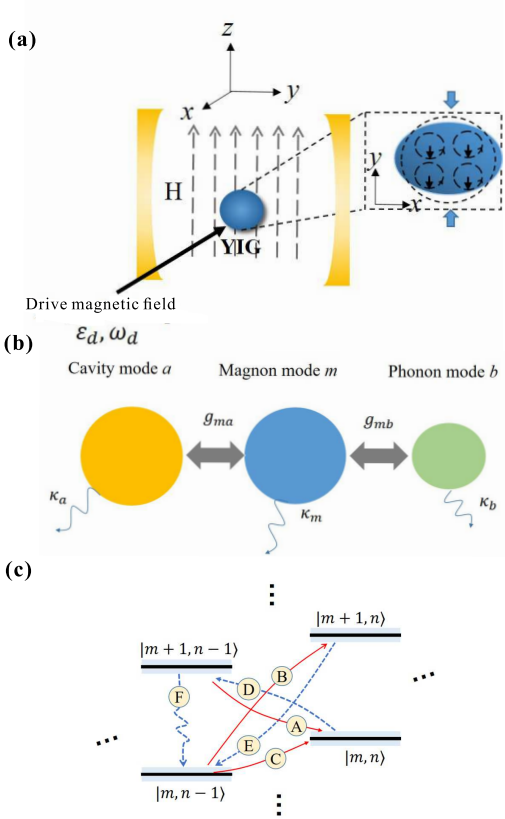} 
\caption{(a) Schematic illustration of the system, a YIG sphere is placed in
the maximum magnetic field of a microwave cavity mode. And an uniform
external bias magnetic field H is applied along the z-direction to bias the
YIG sphere. The enlarged YIG sphere on the right illustrates how the dynamic
magnetization of magnon (vertical black arrows) causes the deformation
(compression along the y-direction) of the YIG sphere (and vice versa),
which rotates at the magnon frequency. Furthermore, a microwave source is
used to drive the magnon mode. It's important to note that the bias magnetic
field (z direction), the drive magnetic field (y direction), and the
magnetic field (x direction) of the cavity mode are mutually perpendicular
at the site of the YIG sphere. (b) The equivalent coupled-harmonic-resonator
model. (c) Energy levels diagram of the effective Hamiltonian in Eq.(\protect
\ref{eq13}), $\left\vert m,n\right\rangle $ denotes the number state with $m$
the magnon number of the magnon mode and $n$ the phonon number of the
mechanical mode.}
\end{figure}

The quantum Langevin equations of the linearized Hamiltonian $H_{lin}$ in
Eq.(\ref{eq02}) are given by

\begin{eqnarray}
\dot{a} &=&(i\Delta _{a}-\kappa _{a})a-ig_{ma}^{\ast }m-\sqrt{2\kappa _{a}}%
a_{in},  \notag \\
\dot{m} &=&(i\Delta _{m}-\kappa _{m})m-ig_{ma}a-iG(b+b^{\dagger })
\label{eq6} \\
&&-\sqrt{2\kappa _{m}}m_{in},  \notag \\
\dot{b} &=&(-i\omega _{b}-\gamma _{b})b-i(G^{\ast }m+Gm^{\dagger })-\sqrt{%
2\gamma _{b}}b_{in},  \notag
\end{eqnarray}%
where $a_{in},m_{in}$ and $b_{in}$ are the corresponding noise operators,
the correlation functions can be found in the appendix A. To make the
following result within experimental realizations, the parameters are in
accord with recent cavity magnomechanical work \cite{k1,k0,k151}, i.e., $%
\omega _{a}/2\pi =$ $\omega _{m}/2\pi =10.1GHz$ of the Kittle mode, $\gamma
_{b}/2\pi =100Hz,\kappa _{a}/2\pi =1GHz,\kappa _{m}/2\pi =0.15MHz,\omega
_{b}/2\pi =10MHz$, the coupling strength $g_{ma}/2\pi =20MHz$\ and $%
g_{mb}/2\pi =$ $0.1Hz$. Here, our research is in resolved sideband regime ($%
\kappa _{m}/\omega _{b}<1$).

For this system, cavity mode $a$ does not directly interact with the
mechanical mode $b$. In order to study the influence of magnon mode on
cooling of the mechanical resonator and make our calculations more
convenient, we show a effective Hamiltonian by assuming the mode $a$ is
highly dissipative. In the limit that $\kappa _{a}\gg g_{ma},G,\kappa _{m}$,
we can adiabatically eliminate mode $a$ \cite{g1}. The equation about $a$ in
Eq.(\ref{eq6}) can be solved.

\begin{equation}
a=\frac{ig_{ma}^{\ast }m+\sqrt{2\kappa _{a}}a_{in}}{i\Delta _{a}-\kappa _{a}}%
.  \label{eq8}
\end{equation}

Then we insert Eq.(\ref{eq8}) into the term of $m$ in Eq.(\ref{eq6}), the
effective Langevin equation for $m$ is given by

\begin{equation}
\dot{m}=(i\Delta _{eff}-\kappa _{eff})m-iG(b+b^{\dagger })-m_{eff,in},
\label{eq9}
\end{equation}%
where the effective parameters can be written as

\begin{equation}
\Delta _{eff}=\Delta _{m}-\frac{\left\vert g_{ma}\right\vert ^{2}}{\Delta
_{a}^{2}+\kappa _{a}^{2}}\Delta _{a},  \label{eq10}
\end{equation}

\begin{equation}
\kappa _{eff}=\kappa _{m}+\frac{\left\vert g_{ma}\right\vert ^{2}}{\Delta
_{a}^{2}+\kappa _{a}^{2}}\kappa _{a},  \label{eq11}
\end{equation}

\begin{equation}
m_{eff,in}=\frac{ig_{ma}\sqrt{2\kappa _{a}}a_{in}}{i\Delta _{a}-\kappa _{a}}-%
\sqrt{2\kappa _{m}}m_{in},  \label{eq12}
\end{equation}%
where $\Delta _{eff}$ is the effective detuning between input drive magnetic
field and magnetic resonance, $\kappa _{eff}$ is the effective decay rate of
the magnon mode and $m_{eff,in}$ is the effective noise operator of the
magnon mode. Note that the effective coupling strength is still $G.$

After adiabatically eliminating optical mode $a$ and using the effective
parameters we got before, the effective Hamiltonian reads

\begin{equation}
H_{eff}=-\Delta _{eff}m^{\dagger }m+\omega _{b}b^{\dagger }b+(Gm^{\dagger
}+G^{\ast }m)(b+b^{\dagger }).  \label{eq13}
\end{equation}

This effective Hamiltonian make the system transform from the three-level
system to an effective two-level system. The above $H_{eff}$ can also be
obtained by the method of effective master equation \cite{g8}.

In Fig. 1(c), we show the level diagram of the linearized Hamiltonian in Eq.(%
\ref{eq13}). There are three kinds of heating processes corresponding to A,
B and C. A represents the swap heating, B represents the quantum backaction
heating and C represents thermal heating, C is an incoherent process arising
from the interaction between the mechanical object and the environment. Our
ultimate goal is to suppress thermal heating, the swap heating and quantum
backaction heating are the accompanying effect, corresponding to the
coherent interaction processes $mb^{\dagger }$ and $m^{\dagger }b^{\dagger }$%
, respectively. The swap heating originates from the energy exchange between
the magnon mode and the phonon mode, it emerges when the system is in the
strong magnomechanical coupling regime. Meanwhile, quantum backaction
heating can lead to a fundamental limit for backaction cooling. The cooling
processes D, E, and F are related to the energy swapping,
counter-rotating-wave interaction between the magnon mode and the phonon
mode, and the dissipation of magnon mode.

\begin{figure}[tbp]
\centering\includegraphics[width=8.2cm]{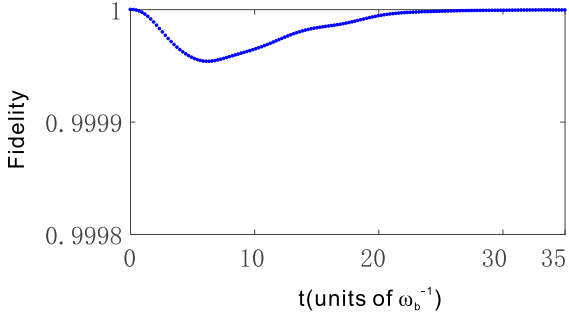} 
\caption{The numerical result of fidelity between the exact state $\protect%
\rho $ and the effective state $\protect\rho _{eff}$ as a function of $t(%
\protect\omega _{b}^{-1}).$ The parameters we chosen are the same as those
in Fig. 3.}
\end{figure}

In order to better prove the accuracy of $H_{eff}$ obtained, we show the
fidelity between the exact state $\rho $ and the effective state $\rho
_{eff} $ as a function of $t(\omega _{b}^{-1})$ in Fig. 2, where $\rho $ and
$\rho _{eff}$ are the numerical results of $H_{lin}$ in Eq.(\ref{eq02}) and $%
H_{eff}$ in Eq.(\ref{eq13}) calculated by quantum master equation $\dot{\rho}%
=i[\rho ,H]+\kappa _{a}\mathcal{D}[a]\rho +\kappa _{m}\mathcal{D}[m]\rho
+\gamma _{b}(n_{th}+1)\mathcal{D}[b]\rho +\gamma _{b}n_{th}\mathcal{D}%
[b^{\dagger }]\rho $, where $\mathcal{D}[o]\rho =o\rho o^{\dagger
}-o^{\dagger }o\rho /2-\rho o^{\dagger }o/2$ $(o=a,m,b)$ are Lindblad
superoperators of our system. Here we choose the same initial state, so
fidelity evolved from 1, then it decreases slightly over time and reaches $1$
finally. The reason for slight decrease is the cavity mode $a$ is highly
dissipative. Fig. 2 implies that our effective Hamiltonian $H_{eff}$ is
valid and a good approximation.

\section{Magnomechanical Cooling}

We analyze the ground-state cooling of the whole system in the cases of weak
and strong magnomechanical coupling. In the weak magnomechanical coupling
regime, similar to dealing with such problems in optomechanical systems, we
use quantum noise approach to get the extra magnetic damping $\Gamma
_{_{md}} $ for the magnomechanical resonator (see Appendix B)%
\begin{eqnarray}
\Gamma _{md} &=&-2\text{Im}\Sigma (\omega _{b})  \label{eq03} \\
&=&2\left\vert G\right\vert ^{2}\text{Re}[\frac{1}{-i(\omega _{b}+\Delta
_{eff})+\kappa _{eff}}  \notag \\
&&-\frac{1}{-i(\omega _{b}-\Delta _{eff})+\kappa _{eff}}].  \notag
\end{eqnarray}

By solving the rate equation in the steady state, the final phonon number
can be analytically described by

\begin{equation}
n_{f}=\frac{A_{+}+\gamma _{b}n_{th}}{\Gamma _{_{md}}+\gamma _{b}},
\label{eq04}
\end{equation}

\begin{figure}[tbp]
\centering\includegraphics[width=7.5cm]{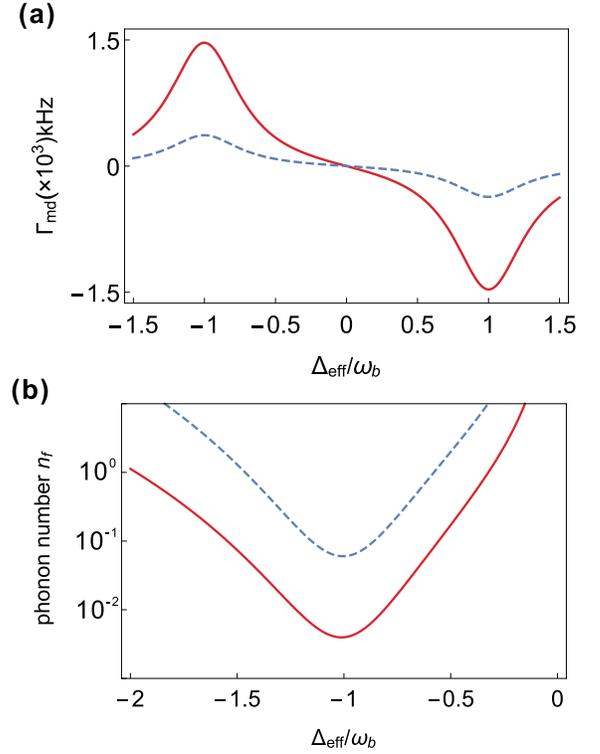} 
\caption{(a) damping $\Gamma _{_{md}}$ versus $\Delta _{eff}$ with two
coupling strengths $G/\protect\omega _{b}=0.15$ (red solid curve) and $0.075$
(blue dotted curve). (b) the final phonon number $n_{f}$ versus $\Delta
_{eff}$ with two coupling strengths $G/\protect\omega _{b}=0.15$ (red solid
curve) and $0.075$ (blue dotted curve). Here we set $\protect\omega _{b}/2%
\protect\pi =10MHz$, $\protect\gamma _{b}/\protect\omega %
_{b}=10^{-5},n_{th}=1000,\protect\kappa _{eff}/\protect\omega _{b}\simeq 0.3$%
, $\protect\kappa _{m}/\protect\omega _{b}=0.15,\protect\kappa _{a}/\protect%
\omega _{b}=1\times 10^{2},$ $g_{ma}/\protect\omega _{b}=2$ and $\Delta _{a}/%
\protect\omega _{b}=1.$}
\end{figure}
where $A_{+}$ is the heating rate, its detailed expressions can be found in
Appendix B. $n_{th}=(e^{\hbar \omega _{b}/k_{B}T}-1)$ is thermal phonon
number of the mechanical resonator, $T$ is the environmental temperature and
$k_{B}$ is the Boltzmann constant. Here, $n_{f}$ can also be regarded as the
cooling limit. And it can be divided into two parts, one part is the
classical cooling limit $\gamma _{b}n_{th}/(\Gamma _{_{md}}+\gamma _{b})$,
the other part is the quantum cooling part $A_{+}/(\Gamma _{_{md}}+\gamma
_{b})$ and it corresponds to the heating rate $A_{+}$ originates from the
quantum backaction.

Fig. 3(a) shows the extra magnetic damping $\Gamma _{_{md}}$ under different
$\Delta _{eff}/\omega _{b}$ with two coupling strengths $G/\omega _{b}=0.15$
and $G/\omega _{b}=0.075$, respectively. It can be seen that the maximum
magnetic damping (gain) is located at the point $\Delta _{eff}=-\omega
_{b}(\omega _{b}).$ From Eq.(\ref{eq03}) and Eq.(\ref{eq04}), $n_{f}$
decreases with the increase of $\Gamma _{_{md}}$. Fig. 3(b) shows the final
phonon number $n_{f}$ under different $\Delta _{eff}/\omega _{b}$ with two
coupling strengths, respectively. It shows the minimum $n_{f}$ is located at
the detuning point $\Delta _{eff}=-\omega _{b}.$When $G/\omega _{b}=0.15$
and $G/\omega _{b}=0.075$, the minimums of $n_{f}$ are about $10^{-2}$ and $%
10^{-1}$ respectively, ground-state cooling of the magnomechanical resonator
can be achieved. Moreover, compared to the two different $G$ in Fig. 3(a)
and Fig. 3(b), it can be known that the stronger $G$, the better effect of
ground-state cooling in weak coupling regime ($G<$ $\kappa _{eff}$).

\begin{figure}[tbp]
\centering\includegraphics[width=7cm]{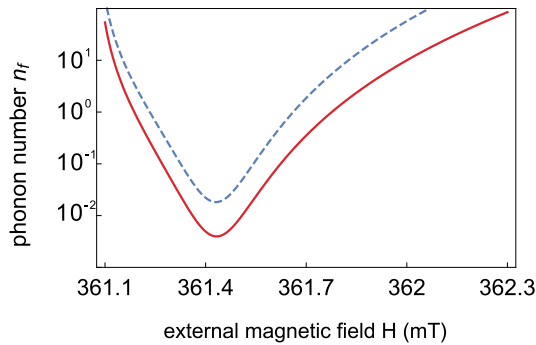} 
\caption{The final phonon number $n_{f}$ versus the external magnetic field $%
H$ with two coupling strengths $G/\protect\omega _{b}=0.15$ (red solid
curve) and $0.1$ (blue dotted curve). The other parameters are the same as
those in Fig. 3.}
\end{figure}

In order to study the influence of magnon mode on the cooling of the
oscillator, in Fig. 4, we plot the final phonon number $n_{f}$ under
different external magnetic field $H$ with two coupling strengths. We find
that for external magnetic field $H$, there is an obvious window in which
the ground-state cooling can appear. It shows the ground-state cooling of
the mechanical resonator can be controlled by tuning $H$, which presents an
additional degree of freedom in cavity optomechanical systems. By comparing
two different coupling strengths $G$, increasing coupling strength widens
the window for ground-state cooling. Note that the external magnetic field $%
H $, the drive magnetic field and the magnetic field of the cavity mode are
mutually perpendicular at the site of the YIG sphere. Consequently, we can
just tune $H$ without affecting the other two.

The recent work shows the strong coupling between magnon mode $m$ and
mechanical mode $b$ can be achieved. From Eq.(\ref{eq13}), the covariance
approach is used to calculate the mean phonon number $N_{bs}$ in steady
state (see Appendix C), and this approach is applicable to both cases of
strong couling and weak coupling.

\begin{eqnarray}
N_{bs} &\simeq &\frac{4\left\vert G\right\vert ^{2}+\kappa _{eff}^{2}}{%
4\left\vert G\right\vert ^{2}(\gamma _{b}+\kappa _{eff})}\gamma _{b}n_{th}
\label{eq05} \\
&&+\frac{4\omega _{b}^{2}(\kappa _{eff}^{2}+8\left\vert G\right\vert
^{2})+\kappa _{eff}^{2}(\kappa _{eff}^{2}-8\left\vert G\right\vert ^{2})}{%
16\omega _{b}^{2}(4\omega _{b}^{2}+\kappa _{eff}^{2}-16\left\vert
G\right\vert ^{2})}.  \notag
\end{eqnarray}

For $G/\omega _{b}=0.4$ $(G>\kappa _{eff}),$ the other parameters are the
same as those in Fig. 3, we have $N_{bs}\simeq 0.257$ by Eq.(\ref{eq05}).
Under the same parameters, the numerical solution of the mean phonon number
obtained from the quantum master equation is shown in Fig. 5. Here, we only
need to numerically calculate the mean values of all the second-order
moments instead of the matrix elements of the density operator $\rho $ (see
Appendix C). The cut-off of the density matrix is not necessary and the
solutions are exact. It can be seen that the numerical solution $N_{bs}$
tends to be stable at $0.28$ after a period of oscillation. This agrees
roughly with the analytical result. Note that the analytical and numerical
solution are both obtained under condition $\Delta _{eff}=-$ $\omega _{b}.$%
The results show that ground-state cooling can be achieved in strong
coupling regime.

The dynamical stability condition of our system can be given by the
Routh-Hurwitz criterion \cite{g7}. Y. C. Liu, $et$ $al$ discussed a similar
two-mode coupled system \cite{g6}. Referring to their results, the
steady-state condition here is $\left\vert G\right\vert ^{2}<\omega
_{b}^{2}/4+\kappa _{eff}^{2}/16$ with the detuning $\Delta _{eff}=-\omega
_{b}$. And the parameters used here are all satisfied with the stability
condition.
\begin{figure*}[tbp]
\centering\includegraphics[width=12cm]{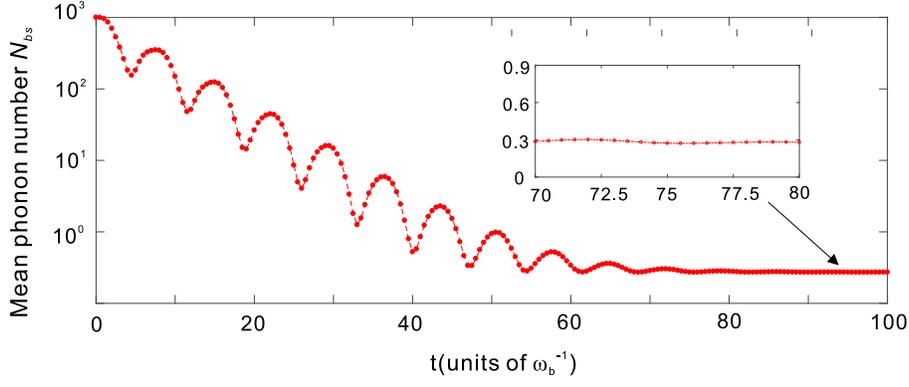} 
\caption{The numerical result of the mean phonon number $N_{bs}$ as a
function of $t(\protect\omega _{b}^{-1}).$ Here, we set $G/\protect\omega %
_{b}=0.15$, the other parameters are the same as those in Fig. 3.}
\end{figure*}
\

Our discussion is on the premise of low-lying excitation, which is $%
\left\langle m^{\dagger }m\right\rangle $ $\ll 2Ns=5N$ ($N$ is the total
number of spins). For a 1-$mm$-diam YIG sphere \cite{k2,k6}, $N=$ $\Lambda
V\approx 2.2\times 10^{18}$, where $\Lambda =4.22\times 10^{27}$ is the spin
density of the YIG and $V$ is the volume of the sphere. For $G/2\pi =4MHz$,
we have $\left\vert \eta \right\vert =4\times 10^{7}$. Finally $\left\langle
m^{\dagger }m\right\rangle =1.6\times 10^{15}\ll 5N\approx $ $1.1\times
10^{19}$, the condition of low-lying excitation is satisfied.

The strong magnon drive we used will led to the unwanted Kerr nonlinear term
$\mathcal{K}m^{\dagger }mm^{\dagger }m$ in the Hamiltonian, where $\mathcal{K%
}$ is the Kerr coefficient. For the 1-$mm$-diam YIG sphere, $\mathcal{K}%
/2\pi \approx 10^{-10}Hz$. For $G/2\pi =4MHz$, we have $\varepsilon
_{d}=4\times 10^{14}Hz.$ When the parameters satisfie the condition $%
\mathcal{K}\left\vert \left\langle m\right\rangle \right\vert ^{3}\ll
\varepsilon _{d}$, Kerr nonlinear term can be neglected. Here, $\mathcal{K}%
\left\vert \left\langle m\right\rangle \right\vert ^{3}=6.4\times 10^{12}\ll
\varepsilon _{d}=4\times 10^{14}Hz$. Therefore, it implies that our
linearized model is a good approximation.

\section{Conclusions}

In summary, we have studied ground-state cooling in a cavity magnomechanical
system, which has three modes: cavity photon mode, magnon mode and phonon
mode. The magnon and mechanical modes are coupled to each other through the
magnetostrictive interaction. By assuming the cavity mode is highly
dissipative, we adiabatically eliminate cavity mode, and the effective
Hamiltonian is given. Which is consist of two mode: magnon mode and phonon
mode. Then we study the final phonon number numerically and analytically.
And we find the ground-state cooling of magnomechanical resonator can be
achieved by using experimentally feasible parameters. Different from the
existing optomechanical cooling system, the extra magnetic damping is the
reason of cavity magnomechanical cooling intrinsically. In other words, we
can utilize magnon mode to achieve the cooling of mechanical mode.
Furthermore, the ground-state cooling can be well controlled by adjusting
the magnetic field without changing other parameters, which provides an
additional degree of freedom.

Because the cavity magnomechanical system has intrinsic great tunability,
low loss, and promising integration with electromechanical elements, we
believe that the proposed scheme provides a promising platform to the
further investigation of cooling of mechanical resonator. And we hope it
opens up new way to the foundations of quantum physics and applications. In
addition, cooling the mechanical system to its quantum ground state is also
an important guarantee for realizing quantum operations in quantum
information processing.

\section*{ACKNOWLEDGEMENTS}

We thank Y. X Zeng for his fruitful discussion. This work was supported by
National Natural Science Foundation of China (NSFC): Grants Nos. 11574041
and 11375036.

\appendix

\section{linearization of Hamiltonian\newline
}

From Eq.(\ref{eq01}), the quantum Langevin equations (QLEs) of the system
are given by

\begin{subequations}
\begin{eqnarray}
\dot{a} &=&(i\Delta _{a}-\kappa _{a})a-ig_{ma}m-\sqrt{2\kappa _{a}}a_{in},
\notag \\
\dot{m} &=&(i\Delta _{m}-\kappa _{m})m-ig_{ma}a-ig_{mb}m(b+b^{\dagger })
\label{eq2} \\
&&+\varepsilon _{d}-\sqrt{2\kappa _{m}}m_{in},  \notag \\
\dot{b} &=&(-i\omega _{b}-\gamma _{b})b-ig_{mb}m^{\dagger }m-\sqrt{2\gamma
_{b}}b_{in},  \notag
\end{eqnarray}
where $a_{in},m_{in}$ and $b_{in}$ are the corresponding noise operators
with zero mean values, and the correlation functions for these noise
operators can be written as
\end{subequations}
\begin{subequations}
\begin{eqnarray}
\left\langle a_{in}(t)a_{in}^{\dagger }(t^{\prime })\right\rangle &=&\delta
(t-t^{\prime }),  \notag \\
\left\langle a_{in}^{\dagger }(t)a_{in}(t^{\prime })\right\rangle &=&0,
\notag \\
\left\langle m_{in}(t)m_{in}^{\dagger }(t^{\prime })\right\rangle &=&\delta
(t-t^{\prime }),  \label{eq201} \\
\left\langle m_{in}^{\dagger }(t)m_{in}(t^{\prime })\right\rangle &=&0,
\notag \\
\left\langle b_{in}(t)b_{in}^{\dagger }(t^{\prime })\right\rangle
&=&(n_{th}+1)\delta (t-t^{\prime }),  \notag \\
\left\langle b_{in}^{\dagger }(t)b_{in}(t^{\prime })\right\rangle
&=&n_{th}\delta (t-t^{\prime }),  \notag
\end{eqnarray}
where $n_{th}$ is thermal phonon number of the mechanical resonator, and it
can be regarded as $n_{th}=(e^{\hbar \omega _{b}/k_{B}T}-1)$, $T$ is the
environmental temperature and $k_{B}$ is the Boltzmann constant.

Then we rewrite each Heisenberg operator as a sum of its steady-state mean
value and the quantum fluctuations, i.e., $a=\alpha +a,m=\eta +m$ and $%
b=\beta +b.$ By separating the classical and quantum components, the quantum
Langevin equations (QLEs) can be rewritten as
\end{subequations}
\begin{subequations}
\begin{eqnarray}
\dot{a} &=&(i\Delta _{a}-\kappa _{a})a-ig_{ma}^{\ast }m-\sqrt{2\kappa _{a}}%
a_{in},  \notag \\
\dot{m} &=&(i\tilde{\Delta}_{m}-\kappa _{m})m-ig_{mb}\eta (b+b^{\dagger })
\label{eq3} \\
&&-ig_{mb}m(b+b^{\dagger })-ig_{ma}a-\sqrt{2\kappa _{m}}m_{in},  \notag \\
\dot{b} &=&(-i\omega _{b}-\gamma _{b})b-ig_{mb}(\eta ^{\ast }m+\eta
m^{\dagger })-ig_{mb}m^{\dagger }m  \notag \\
&&-\sqrt{2\gamma _{b}}b_{in},  \notag
\end{eqnarray}%
where $\tilde{\Delta}_{m}=\Delta _{m}-g_{mb}(\beta +\beta ^{\ast })$. Here,
under the strong driving condition, the nonlinear terms $ig_{mb}m(b+b^{%
\dagger })$ and $ig_{mb}m^{\dagger }m$ can be neglected, then we have
linearized quantum Langevin equations, and the Hamiltonian in Eq.(\ref{eq01}%
) is rewritten as an linearized Hamiltonian
\end{subequations}
\begin{subequations}
\begin{eqnarray}
H_{lin} &=&-\Delta _{a}a^{\dagger }a-\tilde{\Delta}_{m}m^{\dagger }m+\omega
_{b}b^{\dagger }b  \notag \\
&&+(Gm^{\dagger }+G^{\ast }m)(b+b^{\dagger })  \notag \\
&&+(g_{ma}m^{\dagger }a+g_{ma}^{\ast }ma^{\dagger }),  \label{eq5}
\end{eqnarray}%
where $G=\eta g_{mb}$ is the coherent-driving-enhanced magnomechanical
coupling strength, and $\tilde{\Delta}_{m}=\Delta _{m}-g_{mb}(\beta +\beta
^{\ast })$ is the modified detuning of the magnon mode.

\section{weak coupling}

Similar to the method used in cavity optomechanical systems, we study the
cooling rate of the magnomechanical resonator by using the quantum noise
spectrum of the magnetic force.

In weak coupling regime ($G<$ $\kappa _{eff}$), the quantum noise approach
is feasible. From Eq.(\ref{eq13}), $F_{m}(t)=\tilde{m}\ddot{x}+\tilde{m}%
\omega _{b}^{2}x$ and $\dot{p}+\tilde{m}\omega _{b}^{2}x=-\frac{1}{x_{ZPF}}%
[G^{\ast }m(t)+Gm^{\dagger }(t)]$, the magnetic force operator can be
obtained as

\end{subequations}
\begin{subequations}
\begin{eqnarray}
F_{m}(t)=-\frac{1}{x_{ZPF}}[G^{\ast }m(t)+Gm^{\dagger }(t)],  \label{eq14}
\end{eqnarray}
where $\tilde{m}$ is the effective mass of the mechanical resonator, $p$ is
the momentum operator, $x=x_{ZPF}(b+b^{\dagger })$ is the position operator
and $x_{ZPF}=\sqrt{1/(2\tilde{m}\omega _{b})}$($x_{ZPF}$ is the zero-point
fluctuation amplitude of the mechanical oscillator). Using the Fourier
transform of the autocorrelation functions, the quantum noise spectrum is
given by

\end{subequations}
\begin{subequations}
\begin{eqnarray}
S_{FF}(\omega )=\int \left\langle F_{m}(t)F_{m}(0)\right\rangle e^{i\omega
t}dt.  \label{eq15}
\end{eqnarray}

In the absence of the magnomechanical coupling, the term of $m$ in the
Langevin equations of the effective Hamiltonian $H_{eff}$ can be expressed as

\end{subequations}
\begin{subequations}
\begin{eqnarray}
-i\omega m(\omega ) &=&(i\Delta _{eff}-\kappa _{eff})m(\omega )  \notag \\
&&-\sqrt{2\kappa _{eff}}m_{eff,in}.
\end{eqnarray}
\ \

Here, we transform $m(t)$ to the frequency domain. Then using Eq.(\ref{eq14}%
) and Eq.(\ref{eq15}), the spectral density of the magnetic force is
described by

\end{subequations}
\begin{subequations}
\begin{equation*}
S_{FF}(\omega )=\frac{2\kappa _{eff}\left\vert G\chi (\omega )\right\vert
^{2}}{x_{ZPF}^{2}},
\end{equation*}%
where the response function is

\end{subequations}
\begin{subequations}
\begin{eqnarray}
\chi (\omega )=\frac{1}{-i(\omega +\Delta _{eff})+\kappa _{eff}}.
\label{eq18}
\end{eqnarray}
\

The cooling and heating rates can be given by $A_{-}=S_{FF}(\omega
)x_{ZPF}^{2}$ and $A_{+}=S_{FF}(-\omega )x_{ZPF}^{2}$, respectively. And
they correspond to the rates for absorbing and emitting a phonon,
respectively.

By considering the magnomechanical coupling, the Langevin equations of the
effective Hamiltonian $H_{eff}$ are given by
\end{subequations}
\begin{subequations}
\begin{eqnarray}
-i\omega m(\omega ) &=&(i\Delta _{eff}-\kappa _{eff})m(\omega )  \notag \\
&&-iG[b(\omega )+b^{\dagger }(\omega )]  \notag \\
&&-\sqrt{2\kappa _{eff}}m_{eff,in}(\omega ), \\
-i\omega b(\omega ) &=&(-i\omega _{b}-\gamma _{b})b(\omega )  \notag \\
&&-i[G^{\ast }m(\omega )+G(\omega )m^{\dagger }(\omega )]  \notag \\
&&-\sqrt{2\gamma _{b}}b_{in}(\omega ),
\end{eqnarray}
where noise operators $m_{eff,in}(\omega )$ and $b_{in}(\omega )$ are the
corresponding noise operators in the frequency domain. from which we obtain

\end{subequations}
\begin{subequations}
\begin{equation*}
b(\omega )\simeq \frac{\sqrt{2\gamma _{b}}b_{in}(\omega )-i\sqrt{2\kappa
_{eff}}f(\omega )}{i\omega -i[\omega _{b}+\Sigma (\omega )]-\gamma _{b}},
\end{equation*}%
where $f(\omega )=G^{\ast }\chi (\omega )m_{eff,in}(\omega )+G\chi ^{\ast
}(-\omega )m_{eff,in}^{\dagger }(\omega )$. The reason for the approximation
is we consider $\omega \simeq \omega _{b}$, in this way, the terms
containing $b^{\dagger }(\omega )$ can be neglected. $\Sigma (\omega
)=-i\left\vert G\right\vert ^{2}[\chi (\omega )-\chi ^{\ast }(-\omega )]$ is
the magnomechanical self energy. Then the frequency shift $\delta \omega
_{b} $ and the extra magnetic damping $\Gamma _{_{md}}$ are given as

\end{subequations}
\begin{subequations}
\begin{eqnarray}
\delta \omega _{b} &=&\text{Re}\Sigma (\omega ),  \label{eq22} \\
\Gamma _{_{md}} &=&-2\text{Im}\Sigma (\omega )=A_{-}-A_{+}.  \label{eq23}
\end{eqnarray}

Using similar methods in cavity optomechanical systems, from the rate
equation for the probability $P_{n}(t)$ and the average phonon number $\bar{n%
}=\sum_{n=0}^{\infty }nP_{n}$ \cite{k00}, and after making \bigskip $\dot{%
\bar{n}}=0$, we have the final phonon number $n_{f}$

\end{subequations}
\begin{subequations}
\begin{eqnarray}
n_{f}=\frac{A_{+}+\gamma _{b}n_{th}}{\Gamma _{_{md}}+\gamma _{b}}.
\label{eq24}
\end{eqnarray}

\section{Strong coupling}

Because of the introduction of external drive magnetic field, the whole
system can enable strong coupling ($G>$ $\kappa _{eff}$) between magnon mode
$m$ and mechanical mode $b$ by adjusting the intensity of driving field.

For the linear regime under strong driving, the mean phonon number can be
computed exactly by the quantum master equation. From the effective
Hamiltonian in Eq.(\ref{eq13}), the quantum master equation can be described
by

\end{subequations}
\begin{subequations}
\begin{eqnarray}
\dot{\rho} &=&i[\rho ,H_{eff}]m+\kappa _{eff}\mathcal{L(}m)\rho  \notag \\
&&+\gamma _{b}(n_{th}+1)\mathcal{L(}b)\rho  \notag \\
&&+\gamma _{b}n_{th}\mathcal{L(}b^{\dagger })\rho ,
\end{eqnarray}%
where the Lindblad superoperators are given by

\end{subequations}
\begin{subequations}
\begin{eqnarray}
\mathcal{L(}o)\rho =o\rho o^{\dagger }-o^{\dagger }o\rho /2-\rho o^{\dagger
}o/2(o=m,b).  \label{eq251}
\end{eqnarray}

Since the Hamiltonian is linear, we can only calculate the mean values of
all the second-order moments, Such as $\left\langle m^{\dagger
}m\right\rangle ,\left\langle b^{\dagger }b\right\rangle ,\left\langle
m^{\dagger }b\right\rangle ,\left\langle mb\right\rangle ,\left\langle
m^{2}\right\rangle ,\left\langle b^{2}\right\rangle $ and the conjugation of
the last four terms. In the stable regime, under the conditions $\Delta
_{eff}=-$ $\omega _{b}$ and $4\left\vert G\right\vert ^{2}/(\gamma
_{b}\kappa _{eff})\gg 1$, $N_{bs}$ is calculated as

\end{subequations}
\begin{subequations}
\begin{eqnarray}
N_{bs} &\simeq &\frac{4\left\vert G\right\vert ^{2}+\kappa _{eff}^{2}}{%
4\left\vert G\right\vert ^{2}(\gamma _{b}+\kappa _{eff})}\gamma _{b}n_{th}
\label{eq26} \\
&&+\frac{4\omega _{b}^{2}(\kappa _{eff}^{2}+8\left\vert G\right\vert
^{2})+\kappa _{eff}^{2}(\kappa _{eff}^{2}-8\left\vert G\right\vert ^{2})}{%
16\omega _{b}^{2}(4\omega _{b}^{2}+\kappa _{eff}^{2}-16\left\vert
G\right\vert ^{2})}.  \notag
\end{eqnarray}

Note that no cut-off of the density matrix is required in this solution, and
it holds for both weak and strong coupling regimes.

\end{subequations}

\end{document}